\documentclass[letterpaper,twocolumn,10pt]{article}
\usepackage{usenix-2020-09}

\usepackage{tikz}
\usepackage{amsmath}

\usepackage{filecontents}

\usepackage{graphicx}
\usepackage[linesnumbered,lined,vlined,ruled,commentsnumbered,noend]{algorithm2e}
\usepackage{textcomp}
\usepackage{xcolor}
\usepackage{listings}
\usepackage{url}
\usepackage{xspace}
\usepackage{multirow}
\usepackage{balance}
\usepackage{tikz}
\usepackage{ctable}
\usepackage{listings}
\usepackage{color}
\usepackage{xcolor}
\usepackage{graphicx}
\usepackage{subcaption}
\usepackage[nobiblatex]{xurl}

\usepackage{multirow}

\usepackage{algpseudocode}
\usepackage{caption}

\usepackage{cleveref}
\usepackage{acronym}
\usepackage{fancybox}
\usepackage{verbatim}
\usepackage[normalem]{ulem}
\usepackage{float}
\usepackage{newfloat}
\usepackage{mdframed}
\usepackage{booktabs}
\usepackage{pifont}
\usepackage{listings}
\usepackage{paralist}
\usepackage{colortbl}
\usepackage{makecell}
\usepackage{booktabs}
\usepackage{authblk}

\acrodef{pdos}[PDoS]{Provenance Denial of Service}
\acrodef{pados}[PADoS]{Provenance Assisted Denial of Service}

\definecolor{red}{RGB}{255,0,0}
\definecolor{green}{RGB}{18,220,168}

\newcommand{\ie}{i.e.,\xspace}
\newcommand{\eg}{e.g.,\xspace}

\newcommand*\circled[1]{\tikz[baseline=(char.base)]{
            \node[shape=circle,draw,inner sep=0.5pt] (char) {#1};}}
\newcommand{\code}[1]{\texttt{#1}}

\newcommand{\toolname}{\textsc{Nodrop}\xspace}

\newcommand{\eat}[1]{}
\acrodef{wp}[WP]{Website Fingerprinting}
\acrodef{apt}[APT]{Advanced Persistent Threats}
\acrodef{lol}[LotL]{Living-Off-The-Land}
\acrodef{ids}[IDS]{Intrusion Detection System}
\acrodef{vae}[VAE]{Variational AutoEncoder}
\acrodef{re}[RE]{reconstruction error}
\acrodef{sv}[SV]{Stableness Value}
\acrodef{as}[AS]{anomaly score}
\acrodef{gas}[GAS]{graph anomaly score}
\acrodef{etw}[ETW]{Event Tracing for Windows}
\acrodef{e3}[E3]{Engagement 3}
\acrodef{e5}[E5]{Engagement 5}
\acrodef{ttp}[TTPs]{Tactics, Techniques, and Procedures}
\acrodef{hsg}[HSG]{High-level Scenario Graph}
\acrodef{nlp}[NLP]{Natural Language Processing}
\acrodef{dg}[DG]{Detection Graph}
\acrodef{poi}[POI]{Point of Interest}
\acrodef{iv}[IV]{Important Value}
\acrodef{sg}[SG]{Suspicious Graph}
\acrodef{mttd}[MTTD]{Mean Time to Detect}
\acrodef{soc}[SOC]{Security Operations Center}
\acrodef{tib}[TIB]{Thread Information Block}

\newlength{\MaxSizeOfLineNumbers}%
\definecolor{keywordcolor}{rgb}{0.8,0.1,0.5}
\definecolor{lightlightgray}{gray}{.96}
\definecolor{lightgray}{gray}{.925}
\definecolor{medlightgray}{gray}{0.7}
\definecolor{medgray}{gray}{0.4}
\definecolor{darkgray}{gray}{0.35}
\definecolor{nearblack}{gray}{0.15}

\crefname{component}{Component}{Components}

\begin{document}

\date{}

\title{Auditing Frameworks Need Resource Isolation: A Systematic Study on the Super Producer Threat to System Auditing and Its Mitigation}

\author[1]{Peng Jiang}
\author[1]{Ruizhe Huang}
\author[1*]{Ding Li}
\author[1]{Yao Guo}
\author[1]{Xiangqun Chen}
\author[2]{Jianhai Luan}
\author[2]{Yuxin Ren}
\author[2]{Xinwei Hu}
\affil[1]{MOE Key Lab of HCST, School of Computer Science, Peking University}
\affil[2]{Huawei Technologies}
\affil[*]{Corresponding author}

\renewcommand\Authands{ and }

\maketitle

\pagestyle{empty}

\begin{abstract}
System auditing is a crucial technique for detecting APT attacks. However, attackers may try to compromise the system auditing frameworks to conceal their malicious activities. In this paper, we present a comprehensive and systematic study of the super producer threat in auditing frameworks, which enables attackers to either corrupt the auditing framework or paralyze the entire system. We analyze that the main cause of the super producer threat is the lack of data isolation in the centralized architecture of existing solutions. To address this threat, we propose a novel auditing framework, \toolname, which isolates provenance data generated by different processes with a threadlet-based architecture design. Our evaluation demonstrates that \toolname can ensure the integrity of the auditing frameworks while achieving an average 6.58\% higher application overhead compared to vanilla Linux and 6.30\% lower application overhead compared to a state-of-the-art commercial auditing framework, Sysdig across eight different hardware configurations.
\end{abstract}

\section{Introduction}

Auditing frameworks, such as Linux Audit~\cite{audit} or Sysdig~\cite{sysdig}, play a vital role in provenance analysis systems for enterprise security. Many companies build Security Operations Centers (SOC)~\cite{soc, asi,carbonblack,decianno2014indicators,Falco,anafcheh2018intrusion,gomez2019improvements} based on auditing frameworks. Moreover, a large body of research work uses auditing frameworks to develop intrusion detection systems~\cite{gui2019progressive,king_backtracking_2003-1,hossain_sleuth_2017,camflow,LPM,hifi,dap,9152771,berlin2015malicious,du2017deeplog,wang2020you,Falco,han2020sigl,hassan2019nodoze}. We expect this topic to remain relevant for both industry and academia due to the amount of related work.

Unsurprisingly, attackers are engaged in compromising auditing frameworks. Recent studies have also shown that attackers compromise the kernel module of auditing frameworks to prevent attack traces from being recorded. Paccagnella \textit{et al.}~\cite{paccagnella2020logging} proposed a race condition attack in which an attacker with root privileges can compromise the kernel module of auditing frameworks to hide their malicious actions. To protect the kernel module of auditing frameworks, researchers have proposed multiple approaches, such as Hardlog~\cite{ahmad2022hardlog}, KennyLoggings~\cite{paccagnella2020logging}, and QuickLog~\cite{281386}. There are also many user-space attacks that mainly involve log tampering ~\cite{bowers2014pillarbox,logtempering2,logtempering3,logtempering4}. Meanwhile, several cryptographic-based approaches have been proposed to secure the user space transmission and storage of logs~\cite{Forwardsecure,syslog-ng,10.1145/3052973.3053034,Custos}.

In addition to the vulnerabilities mentioned above, this paper concentrates on the \textit{super producer threat}. This threat does not require attackers to have root privileges to launch attacks on the kernel module. In essence, the attacker can either disable provenance data collectors or intensify DoS attacks on the system under observation by creating a super producer, a process that produces a large number of system provenance events in a brief span of time, with user privilege.

Specifically, by generating a large amount of provenance data that auditing frameworks cannot process in time with a reasonable amount of system resources, a super producer puts the current auditing framework into the \textit{data integrity vs. efficiency} dilemma. On one hand,  auditing frameworks may adopt the \textit{pro-performance} strategy~\cite{sysdig,audit,lttng}, which drops system events if there is too much provenance data. However, under this strategy, attackers may launch the \textit{\ac{pdos} attack}, in which the attacker uses the super producer to evict the events of malicious behaviors from the event buffers of auditing frameworks.  On the other hand,  auditing frameworks may adopt the \textit{pro-integrity} strategy~\cite{Camquery}, which elastically allocates sufficient resources to ensure that all provenance data can be processed in time. Unfortunately, this strategy may lead to the \textit{\ac{pados} attack}, in which attackers exploit the pro-integrity strategy to degrade the performance of the whole system, amplifying the DoS attack on the server. Notably, the \ac{pados} attack can even break the protection of \code{cgroup}.

The main reason for the super producer threat is that the design of existing auditing frameworks breaks the resource and logic isolation of processes, which is critical for modern OSes to achieve high performance for concurrent tasks. Existing solutions collect system events by intercepting system calls in the OS kernel and then processing the collected events in a centralized user-space collector~\cite{sysdig,lttng,Camquery,audit}. The centralized collector handles provenance data equally regardless of the priority, importance, and resource quota of the processes that generate the provenance data. Therefore, a super producer can generate a massive amount of provenance data that occupies all the processing power of the centralized handler and causes the  ``data integrity vs. efficiency dilemma''.

In this paper, we present a comprehensive analysis of the super producer threat and propose a novel auditing framework, \toolname, that balances the trade-off between ``data integrity and efficiency''. Specifically, \toolname surpasses existing solutions in two aspects. First, it guarantees the integrity of provenance data. \toolname records all provenance data faithfully regardless of the workload. Therefore, a super producer cannot conceal the traces of attacks by generating too many system events, thus mitigating the \ac{pdos} attack. Second, \toolname prevents the super producer from degrading the performance of the whole system, avoiding the \ac{pados} attack.

The key insight of our design is to provide isolation to the provenance data collector so that each process consumes its own resource quota to handle the provenance data generated by itself. The logic behind this design is as follows. Like user-space log events (e.g., log4j events), provenance events reflect the status of the running processes. Thus, provenance data should be considered as the logs of the corresponding processes, instead of the OS. Therefore, each process should spend its own resource quota to handle the provenance data it generates.

By isolating the provenance data of each process, we can naturally mitigate the super producer threat. This way, a process that generates a huge amount of provenance data in a short time can only affect its own performance, since it has a limited resource quota. It also cannot interfere with the processing power of other processes. Therefore, the super producer cannot stop the system from recording provenance events of attacks by overwhelming the auditing frameworks.

The main challenge of isolating provenance data from different processes is to achieve efficiency. We use a \textit{threadlet-based} approach that inserts the provenance data processing logic into the memory of running applications. This approach leverages the process isolation strategy of the OS directly, eliminating the extra overhead of adding a new isolation strategy to the auditing framework. This approach also reduces process scheduling and its associated cache miss costs.

We thoroughly evaluate \toolname with eight different hardware configurations and five baselines.
Our evaluation shows that \toolname faithfully records all provenance data, preventing the \ac{pdos} attack. On the contrary, current pro-performance auditing frameworks (\ie Sysdig) can drop up to 90\% of provenance data while the super producer is running. \toolname can also prevent the \ac{pados} attack. Specifically, \toolname only slows down three popular applications by 4.0\% on average, regardless of the workload generated by the super producer. For comparison, existing pro-integrity auditing frameworks can slow down the applications by up to 59.1\% on average. More importantly, when the super producer increases its workload, the application performance decreases accordingly with existing auditing frameworks. \toolname is also efficient compared with the SOTA pro-performance collector, Sysdig. \toolname has, on average, 6.30\% less application overhead than Sysdig. In summary, our evaluation proves that \toolname can address the super producer threat while incurring lower system overhead than existing auditing frameworks.   

To sum up, this paper makes the following contributions:
\begin{itemize}
    \item To the best of our knowledge, this is the first thorough systematic study on the super producer threat and related attacks to auditing frameworks.
    \item We identify that the root cause of the super producer threat is the lack of resource isolation in the user space component of existing auditing frameworks.
    \item To address the super producer threat, we propose a novel auditing framework \toolname that efficiently isolates resources for provenance data handling by enforcing processes to consume their own resource quota to handle the provenance data generated by themselves.
    \item Extensive experiments demonstrate that \toolname can address the super producer threat, as well as its efficiency.  
\end{itemize}
\textbf{Availability:}  \toolname is available at: \url{https://github.com/PKU-ASAL/NoDrop}
\section{Background}
\label{sec:background}

\begin{table*}[!t]
    \caption{A comparison of auditing frameworks. \emph{* means the collector is implemented by us. Per-core thread means each CPU core has a dedicated processing thread for provenance data. Per-core buffer means each CPU core has a dedicated event buffer.}}
    \label{tab:collectors}
    \centering
    \begin{tabular}{@{}ccccc@{}}
    \toprule
     Name       & Computation isolation & Data isolation       & Synchronization & Strategy   \\ \midrule
    Sysdig~\cite{fangback,sabharwal2020monitoring,salamero2019kubernetes,gantikow2019rule,tien2019kubanomaly,grimmer2019modern}       & Single thread    & Per-core buffer         &   Asynchronous     & pro-performance        \\
    Linux Audit~\cite{wang2019attentional,wang2020you,han2020sigl,hassan2019nodoze,inamforensic} & Single thread   & Single buffer           &     Asynchronous     & pro-performance         \\
    LTTng~\cite{kohyarnejadfard2019system,sultana2012improved,kohyarnejadfard2021anomaly,khreich2017anomaly}   & Single thread     & Per-core buffer                  &    Asynchronous      & pro-performance     \\
    Camflow~\cite{camflow,han2020ndss,pasquier2018ccs}     & Per-core thread    & Per-core buffer    &      Asynchronous     & pro-integrity       \\
    KennyLoggings~\cite{paccagnella2020logging}   & Single thread     & Single buffer                  &  Asynchronous      & pro-performance    \\
    Hardlog~\cite{ahmad2022hardlog}   & Single thread     & Single buffer                   &  Synchronous      & pro-performance    \\
    QuickLog~\cite{281386}       & Single thread     &  Single buffer                &  Asynchronous     & pro-performance    \\
    Sysdig-Camflow*      & Per-core thread     & Per-core buffer                      &    Asynchronous    & pro-integrity         \\
    Sysdig-Integrity*      & Single thread    & Per-core buffer                      &    Synchronous   & pro-integrity         \\
    \bottomrule
    \toolname*      & Per-thread threadlet    & Per-threadlet buffer         &   Synchronous    & performance and integrity        \\
    \bottomrule
    \end{tabular}
    \vspace{-2ex}
\end{table*}

Auditing frameworks are the fundamental part of provenance analysis~\cite{king2003backtracking}, which is a technique that monitors system activities to detect and investigate attacks~\cite{fangback,sabharwal2020monitoring,salamero2019kubernetes,gantikow2019rule}. Popular auditing frameworks include Sysdig~\cite{sysdig}, LTTng~\cite{lttng}, and Linux Audit~\cite{audit}. These three auditing frameworks are the most widely cited in provenance-based detection solutions~\cite{fangback,sabharwal2020monitoring,salamero2019kubernetes,gantikow2019rule,tien2019kubanomaly,grimmer2019modern,wang2019attentional,wang2020you,han2020sigl,hassan2019nodoze,inamforensic,kohyarnejadfard2019system,sultana2012improved,kohyarnejadfard2021anomaly,khreich2017anomaly}. Besides, there are more recent auditing frameworks from academia, including Camflow~\cite{camflow,Camquery}, Hardlog~\cite{ahmad2022hardlog}, KennyLoggings~\cite{paccagnella2020logging}, and QuickLog~\cite{281386}. We thoroughly investigate the auditing frameworks published in industry and academia in recent years. We summarized the auditing frameworks in Table~\ref{tab:collectors}.

The existing auditing framework has a ``centralized'' architecture~\cite{sysdig,audit,lttng,Camquery}, as Figure~\ref{fig:sysdig} shows. This framework intercepts provenance events through a kernel module and digests them based on user-specified rules (\eg sending the provenance data to a remote log server or storing it to local files) in a centralized user-space module, called \texttt{collector}. Current auditing frameworks mainly differ in how they handle massive amounts of provenance data. They adopt two strategies: pro-performance and pro-integrity.

Several solutions, such as Sysdig~\cite{sysdig}, Linux Audit~\cite{audit}, and LTTng~\cite{lttng}, follow the ``pro-performance'' strategy. The rationale behind this strategy is that the auditing framework should minimize the system run-time overhead and maintain the performance of critical services on the monitored host. Current solutions limit the CPU usage of the auditing framework by allowing only \textit{one collector thread}. If this thread cannot process all the provenance events in time, it will either stop receiving events or drop them.

Solutions that adopt the ``pro-integrity'' strategy~\cite{camflow,Camquery} try to allocate enough resources to the auditing framework to handle all the provenance events. For instance, Camflow~\cite{camflow} uses a multi-threaded model that dynamically allocates computational resources based on the provenance data generation speed.
 
\section{The Super Producer Threat}
\label{sec:dilemma}

\begin{figure*}[t!]
\centering \includegraphics[width=0.9\textwidth]{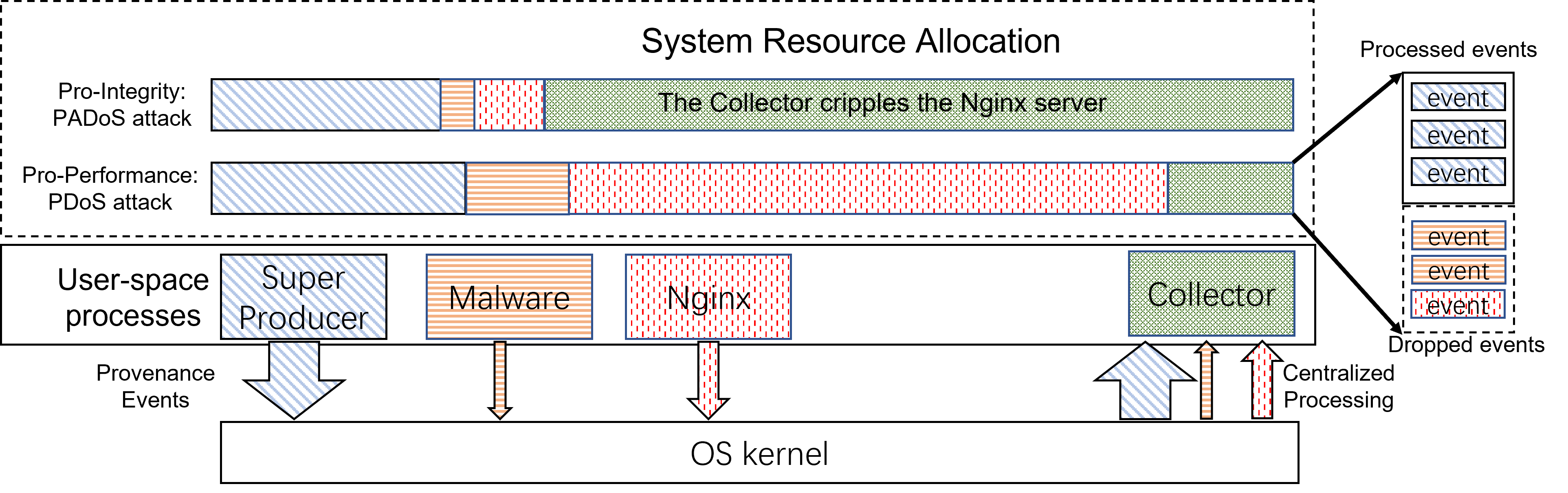}
\vspace{-0.1in}
\caption{\small The design of existing auditing frameworks and the ``data integrity vs. efficiency dilemma''. }
\vspace{-3ex}
\label{fig:sysdig}
\end{figure*}

The existing auditing frameworks use a centralized architecture that exposes them to the super producer threat. This threat occurs when an attacker exploits a super producer to consume the computational resources of other processes, breaking the logic and resource isolation between them. As a result, current auditing frameworks face a dilemma between data integrity and efficiency.

Figure~\ref{fig:sysdig} illustrates the super producer threat and the ``data-integrity vs. efficiency'' dilemma. The figure shows three user-space applications (the super producer, the malware, and the Nginx server) and an auditing framework that processes all the provenance data of these applications. The arrows indicate the direction of provenance data flow, and the width of the arrows reflects the amount of provenance data.

The super producer produces considerable system provenance data that exhausts the collector's computation capacity.
As a result, the collector either drops the provenance data of other applications or competes for more computational resources, implicitly breaking the resource quota of each application.
Thus, it becomes feasible to exploit \ac{pdos} and \ac{pados} attacks.

The pro-performance strategy restricts the resource quota of the collector to prevent performance degradation of the whole system~\cite{sysdig,lttng,audit}, but this exposes the system to the \ac{pdos} attack.  
Figure~\ref{fig:sysdig} shows that the collector's limited resources cannot cope with the high rate of provenance data generation by the super producer, and the collector will drop events when overloaded.
Moreover, the collector does not separate the provenance data from different applications, so other critical provenance events of the malware may be evicted, enabling the attackers to conceal the malware from detection.

The pro-integrity strategy gives the collector more resources to prevent the loss of provenance data~\cite{han2020ndss,pasquier2018ccs}, but this exposes the system to the \ac{pados} attack. 
Figure~\ref{fig:sysdig} shows how the collector consumes more resources to handle all provenance events, while the resources of other applications are reduced accordingly, resulting in significant performance interference for the whole system.
Moreover, since the super producer indirectly affects the system performance by using the collector, it only requires moderate resources to generate a large amount of provenance data.
Hence, existing isolation mechanisms (\eg \code{cgroups}), which limit the resource usage of the super producer, cannot effectively stop the \ac{pados} attack.

\subsection{Research Challenges}
Addressing the super producer threat is conceptually challenging. One possible solution is to suppress the super producer's generation speed of provenance data with some specified threshold. However, this strategy is not systematic. First, it is hard to set an effective threshold considering the dynamics of the systems. Second, attackers may use a set of super-producer processes to avoid reaching the threshold.

Another straightforward solution is to provide isolation inside the user-space collector. However, this strategy requires complex user-space logic in the collector, increasing the run-time overhead and difficulty in adapting to different systems. Note that simply providing a separate event buffer for different processes is not sufficient because other computational resources, such as the CPU, also need to be isolated.
More importantly, resource isolation or scheduling policy inside the collector may be inconsistent or conflict with the original policy made by the OS. Thus, different policies interfere with each other, causing all of them ineffective.

In summary, we need to redesign the auditing framework architecture, which can adaptively suppress the super producer, isolate provenance data, avoid performance interference, and respect the OS resource management policies.

\subsection{Attack Scenario and Threat Model}
\label{sec:case}
We consider a common scenario of a multi-tenant web server~\cite{multitenancy} as the potential context for \ac{pdos} and \ac{pados} attacks. We suppose that two Internet-facing applications, the \textit{target app} and the \textit{victim app}, from different users, are running on the server at the same time. These two applications have their own resource quota (\ie in separate \code{cgroups}). An auditing framework is running to monitor both the \textit{target app} and the \textit{victim app}. For the \ac{pdos} attack, the attacker's objective is to disable the auditing framework by attacking the \textit{victim app}. Then, the attacker can try to compromise the \textit{target app} without leaving traces in the provenance data. For the \ac{pados} attack, the attacker's aim is to paralyze the \textit{target app} by compromising the \textit{victim app}.

We define our threat model as follows: the attacker can transform the \textit{victim app} into a super producer. This does not imply that the attacker has to compromise the \textit{victim app}. It is enough to generate a large number of requests for a complex dynamic web application. Moreover, we assume that the attacker is aware of the auditing framework that is deployed on the server.

We also make the following assumptions about the security of the auditing frameworks. First, the kernel modules that collect provenance data are not vulnerable to attacks. Second, the provenance data is stored and transmitted securely and reliably. Third, the user space module that analyzes the provenance data is protected by existing intrusion detection systems that can alert us if the attacker tries to disable or compromise the auditing frameworks~\cite{sysdig,lttng}. The protection of the kernel module, the transmission, and the storage of provenance data is beyond the scope of this paper.

\section{Design of \toolname}
\label{sec:design}
This section describes the design of \toolname, which aims to prevent \ac{pdos} and \ac{pados} attacks by attacks by achieving thread isolation for provenance data processing. Similar to Sysdig, \toolname captures system call, thread switching, and system signal events, and allows users to provide custom logic for processing these events.

\subsection{Design Goals} 
\toolname is designed to address the super producer threat by satisfying the following properties:

\noindent \textbf{G1 - Zero Data Lost.} \toolname must record all provenance data generated by the system to prevent \ac{pdos} attacks.

\noindent \textbf{G2 - Performance Isolation.} \toolname must ensure that a super producer cannot affect overall system performance, preventing attacks such as \ac{pados}.

\noindent \textbf{G3 - Low Overhead.} \toolname should not introduce significantly higher costs than existing auditing frameworks.

\subsection{Design Principles} \toolname is guided by two design principles: self-consuming execution and synchronized logging buffer.

\noindent \textbf{Self-consuming execution.} With \toolname, each thread processes its own generated provenance data. This differs from the centralized processing architecture of current auditing frameworks and addresses the super producer threat by achieving resource and data isolation. Each thread uses its own resource quota to process its provenance data in a dedicated buffer, preventing a super producer from evicting other threads’ system events or slowing them down by occupying their resource quota.

\noindent \textbf{Synchronized logging buffer.} \toolname allocates a dedicated logging buffer for each thread and dynamically instruments provenance data processing logic into each thread when the buffer is full. This synchronized strategy ensures no provenance data is overwritten or lost, regardless of the super producer’s data volume.

These principles are realized through a threadlet-based approach. A threadlet is a self-contained piece of code that exists in the memory space of a host thread. \toolname instantiates provenance data processing logic as a threadlet, inserts it into the memory of the currently running thread, and consumes the provenance data stored in the dedicated buffer. This type of threadlet is referred to as a consumer throughout the paper.

The threadlet-based design of \toolname offers three advantages: \begin{inparaenum}[(1)] \item It allows \toolname to use each running thread’s resource quota to process the provenance data generated by that thread, without breaking the OS’s resource isolation mechanism or slowing down the entire system. \item Synchronized insertion of a threadlet into a running thread’s memory space provides data isolation, preventing a super producer from manipulating or overwriting provenance data in other threads. \item The threadlet-based architecture is more resource-efficient and lightweight than a conventional thread, eliminating most thread switch overhead such as scheduling delay, cache movement, and priority inversion. \end{inparaenum}

In summary, the threadlet-based architecture satisfies three requirements: \circled{1} all provenance data is collected; \circled{2} a super producer cannot slow down the entire system; and \circled{3} overhead is optimized.

\begin{figure}[!t]
\includegraphics[scale=0.68]{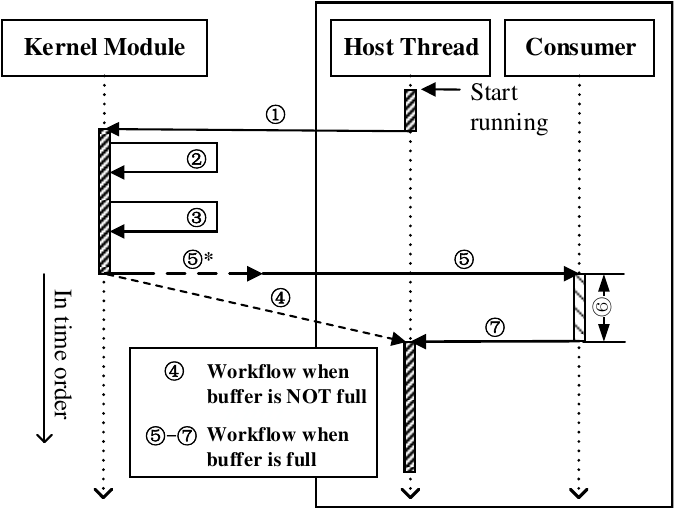}
\caption{The workflow of \toolname. \protect\circled{1} - \protect\circled{7} are the seven steps of \toolname to handle a system call event.} 
\label{fig:workflow}
\vspace{-3ex}
\end{figure}

\subsection{High-Level Workflow}
\toolname consists of two main components: a kernel module and a user-space consumer. Like other auditing frameworks, the kernel module intercepts system calls to collect provenance data and stores it in a dedicated logging buffer. The consumer, implemented as a threadlet, processes the provenance data.
In \toolname, provenance data includes three types of system events: system call, thread switching, and system signals. This definition is the same as that used by Sysdig~\cite{sysdig}.

The workflow of how \toolname handles system calls is shown in Figure~\ref{fig:workflow}. When a thread invokes a system call, the kernel captures it (\circled{1}) and executes it (\circled{2}). The kernel module of \toolname then catches and records the system call in a dedicated in-kernel logging buffer (\circled{3}). If the buffer is not full, \toolname returns control to the host thread (\circled{4}). If the buffer is full or the thread exits, control is passed to the consumer (\circled{5}). Before transferring control, \toolname checks if there is a consumer in the running thread’s memory. If not, it first instruments the consumer into the running thread($\circled{5}^*$). Once the consumer has control, it uses the running thread’s resource quota to process the transferred provenance data, providing performance and data isolation between threads (\circled{6}). When processing is complete, control is returned to the original thread (\circled{7}). The workflow for handling thread switching and system signals differs from that for system calls only in how the kernel is entered (\circled{1} in Figure~\ref{fig:workflow}); all other steps are the same.

\subsection{Design Details}
This section describes the detailed design and implementation of key components in \toolname. Since \toolname’s kernel module is similar to that of Sysdig~\cite{sysdig}, its details are not discussed. Instead, the focus is on the unique design of the in-kernel logging buffer (\S\ref{sssec:buffer}), the user-space consumer (\S\ref{sec:consumer}), and consumer instrumentation (\S\ref{sssec:consumer-init}).

\subsubsection{In-Kernel Logging Buffer}
\label{sssec:buffer}

The in-kernel logging buffer is used to store provenance data in the kernel, improving the efficiency of auditing frameworks. When the buffer is full, the consumer consumes provenance data from the front end. Meanwhile, the kernel module pushes incoming provenance data to the end of the buffer  until the buffer is full.

\toolname uses a per-thread buffering scheme, allocating one logging buffer for each thread. This is chosen to meet design goal G2, which requires isolation of provenance data from each thread. This buffering scheme naturally distinguishes provenance data from every thread.

In \toolname, each system event contains the metadata and the argument data. The metadata includes the basic attributes of a system event, such as its type, timestamp, and size of the system event. The argument data is distinct for different types of system events. For system call events,  each element in the argument data is the value of a system call parameter. If the size of the parameter values is large (\eg the content of \code{read}), \toolname follows the practice of Sysdig by truncating the parameter values that are larger than 80 bytes. For thread switching, the argument data contains the IDs of the previous thread and the next thread, respectively. For signals, the argument data contains the signal ID and the process ID, which captures the signal. 

\subsubsection{The Consumer}
\label{sec:consumer}
The consumer is designed as a user-defined threadlet function for processing provenance data. It has one parameter that points to a copy of the in-kernel logging buffer in user-space and returns nothing. The key task for \toolname is to ensure the efficiency and security of the consumer. To achieve this, the conventional threadlet is improved.

\noindent \textbf{Kernel interaction.}
To efficiently and correctly pass the in-kernel logging buffer to the consumer, \toolname uses \code{mmap} to directly map the buffer from the kernel to user space. This avoids the overhead of copying data and restricts the consumer from accessing other in-kernel memory.

\begin{figure}[]
\centering
\includegraphics[scale=0.55]{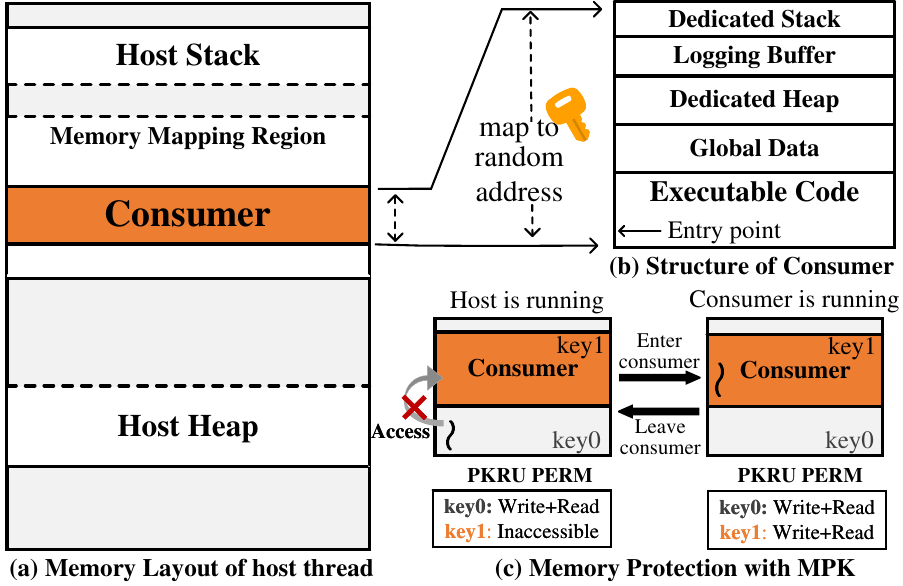}
\caption{\small Memory layout of the host thread (\textit{a}) and structure of consumer (\textit{b}). \emph{The consumer is mapped at a random address in the \code{mmap} region of the host thread. The consumer has its own 5 MPK-protected sections: executable code, global data, dedicated heap plus stack, and the logging buffer mapped from the kernel.}}
\label{fig:layout}
\vspace{-1em}
\end{figure}

\noindent \textbf{Memory protection.}
The key challenge for the consumer is to ensure that it cannot be compromised by the host process. Since the host thread or other parallel threads are in the same process as the consumer, they could potentially modify the consumer’s data. To prevent this, \toolname uses a comprehensive approach that combines address space randomization, an isolated heap, and MPK to protect the consumer. Figure~\ref{fig:layout} summarizes the memory protection mechanisms used by \toolname. The consumer is located in a randomized memory region with a dedicated protection key. This region includes a separate stack, heap, and mapped logging buffer.

First, \toolname randomizes the consumer’s address to prevent attackers from obtaining it, making it more difficult to compromise the consumer. When the consumer is first instrumented, \toolname randomly chooses the threadlet’s loading address, preventing attackers from obtaining the consumer’s address before the threadlet is run.

Second, \toolname allocates a dedicated heap and stack for the consumer to prevent memory-overflow-based attacks (e.g., ROP) from the host thread. Isolating the heap is challenging because an attacker could potentially compromise the heap allocator that allocates memory from the consumer’s dedicated heap region. To prevent this, \toolname pre-loads a customized heap allocator for the consumer that always allocates memory from the dedicated heap.

Finally, \toolname enforces memory isolation using MPK, a hardware primitive that achieves in-process memory isolation by controlling protection keys and their permissions with a thread-local register called \code{PKRU}~\cite{vahldiek2019erim,234966}. To protect the consumer, \toolname binds a dedicated \code{key} to all consumer memory. By updating the \code{PKRU} value, access permission for that key is enabled at the consumer entry point and disabled after it exits. Similar to previous approaches using MPK~\cite{vahldiek2019erim,234966}, \toolname ensures that the host thread cannot modify the \code{PKRU} register by scanning executable code to validate that it contains no \code{PKRU}-related instructions. In \toolname, all consumers in the same process share the same key. Since the total number of MPK keys is limited, enough keys are left for later use. Using one MPK key does not compromise \toolname’s security because MPK protection is thread-local~\cite{234966,mpk-intro}.

\noindent \textbf{Privilege escalation.}
 Sandbox or container threads may have limited execution privileges. 
However, suppose the consumer uses the sandbox or container threads as host threads and inherits the limited privilege. In that case, it cannot perform necessary processing on the provenance data, such as writing the data to files or network~\cite{gui2019progressive,king_backtracking_2003-1,hossain_sleuth_2017,camflow,LPM,hifi,dap}. 
To this end, \toolname elevates the consumer's privileges by raising the \code{RLIMIT} and \code{CAPACITY} to unlimited and disabling the \code{SECCOMP}.
\toolname recovers these privileges when switching back to the sandbox or container thread.

\noindent \textbf{Atomic execution.}
Because of the privilege escalation, the sandbox or container thread may abuse the higher privilege of the consumer, thus breaking the system security policy.
To avoid such abuse and other data-racing between the consumer and the application thread, \toolname ensures the atomicity of the consumer execution by disabling all signals while the consumer is running. \toolname delays the upcoming signals until the consumer exits. 
\toolname enables all signals when switching back to the sandbox or container thread. At this moment, \toolname delivers the delayed signals to their handlers of the host thread.
Therefore, there is never interleaved execution between the consumer and the sandbox or container thread.

\subsubsection{Consumer Instrumentation} 
\label{sssec:consumer-init}

\toolname blocks the running thread and invokes the consumer (\protect\circled{5} in Figure~\ref{fig:workflow}) when either the in-kernel logging buffer is full or the running thread exits. The instrumentation process of the consumer is described in detail here.

\noindent \textbf{Threadlet initialization.}
\toolname prepares and initializes the consumer execution environment when the running thread traps into the kernel for the first time (\protect\circled{1} in Figure~\ref{fig:workflow}).
Specifically, \toolname allocates the space of the in-kernel buffer and a per-consumer control block. 
The control block stores the meta information describing a consumer, including its running state, the address of the user-defined entry function, memory layout, and protection key.

\noindent \textbf{Consumer loading.}
When the consumer is first instrumented, \toolname loads its binary($\protect\circled{5}^*$ in Figure~\ref{fig:workflow}). The loading process is similar to that of the \code{execve} system call. The kernel module reads and parses the consumer’s ELF file, allocates memory space, and loads all segments into memory, as shown in Figure~\ref{fig:layout}(b). Additionally, \toolname reserves an MPK key and initializes the consumer control block. This loading phase only occurs once for each thread, so its overhead does not significantly degrade system performance. Furthermore, existing in-memory template caching techniques can be leveraged to further optimize this cost~\cite{edgeos_atc_20}.

\noindent \textbf{Consumer invocation.}When invoking the consumer (\protect\circled{5} in Figure~\ref{fig:workflow}), \toolname's kernel module maps the logging buffer, saves the register context of the running thread, updates \code{PKRU} value, elevates the privilege and finally upcalls to the consumer's entry point.

\noindent \textbf{Consumer exit.}
We add a new system call that the consumer uses to exit (\protect\circled{7} in Figure~\ref{fig:workflow}).
This system call does the reverse of consumer invocation.
Concretely, it releases the logging buffer, recovers the privilege configuration, restores the \code{PKRU} and other registers, and lastly, switches back to the instrumented thread.

\section{Evaluation}\label{section5}

In this section, we evaluate whether \toolname can address the super producer threat without introducing significantly higher system overhead than existing approaches. Specifically, we focus on the following questions:

\begin{itemize}
    \item RQ 1: Can \toolname avoid dropping provenance data?
    \item RQ 2: Can \toolname prevent a super producer from slowing down other applications?
    \item RQ 3: What is the run-time overhead of \toolname?
    \item RQ 4: Can data reduction techniques address the super-producer threat?
    \item RQ 5: Can increasing the buffer size address the super-producer threat?
\end{itemize}

To ensure the generalizability of our evaluation, we run experiments on four hardware configurations: 1 CPU core with 2GB memory (C1 and C5), 4 CPU cores with 8GB memory (C2 and C6), 16 CPU cores with 32GB memory (C3 and C7), and 32 CPU cores with 64 GB memory (C4 and C8). We also conduct our experiments on both physical and virtual machines, resulting in eight different configurations in total. C1-C4 represent VM configurations, while C5-C8 represent PM configurations. All machines run OSes of Ubuntu 18.04. \toolname is also tested on other Linux distribution like openEuler~\cite{oe} 20.03 which shows the similar results.

\subsection{RQ 1: Event Drop}
\label{ssec:eventdrop}
To answer this research question, we first conducted a controlled measurement study on the number of provenance events dropped by \toolname and pro-performance solutions such as Sysdig, LTTng, and Linux Audit. Then, we simulated a realistic web server to evaluate how well \toolname can prevent \ac{pdos} attacks.

In the controlled experiment, we launched a bash script to mimic a super producer that generates a large amount of provenance data in a short period. The script forks $n$ processes, where $n$ is the number of CPU cores, and each process repeatedly invokes the \code{write} system call and performs the \code{count++} operation. We adjusted the proportion of the \code{write} and \code{count++} operations to control the generation speed of system call events over 30 seconds. We ran the experiment on four hardware configurations, on both virtual and physical machines. The percentage of dropped events represents the potential success rate of a \ac{pdos} attack.

\begin{figure*}
    \centering
    \includegraphics[width=\textwidth]{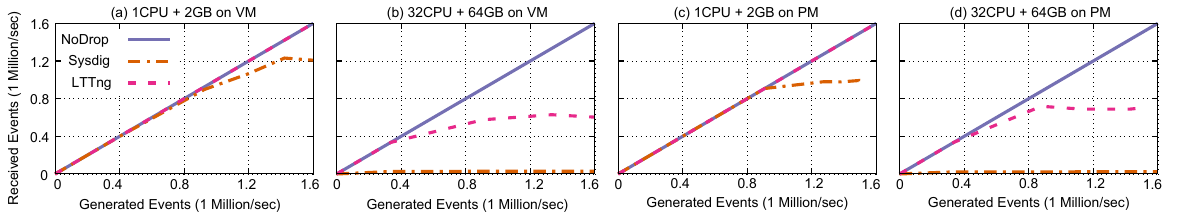}
    \vspace{-0.1in}
    \caption{\small The number of events dropped by \toolname and baselines with different hardware configurations. \emph{The x-axis is the number of generated events in 30 seconds, and the y-axis is the number of events handled by the auditing framework.}}
    \label{fig:droppaper}
    \vspace{-1em}
\end{figure*}

Figure~\ref{fig:droppaper}(a),~\ref{fig:droppaper}(b),~\ref{fig:droppaper}(c), and~\ref{fig:droppaper}(d) show the results for cases with 1 CPU core and 2GB memory, and 32 CPU cores with 64 GB memory, on both virtual and physical machines. The results for other configurations are similar to those shown in Figure~\ref{fig:droppaper}, but due to space limitations, we have included them in our GitHub repository. The x-axis represents the number of system call events generated by the kernel per second, while the y-axis represents the number of system call events processed per second by the user-space component. The gap between y and x represents the number of events dropped. The diagonal line in the figure is the ideal line, indicating that no events have been dropped. The blue dot-dashed line and magenta dash line represent Sysdig and LTTng, respectively. We have omitted the line for Linux Audit for clarity, as it drops nearly all events while the super producer is running.

Our evaluation shows that \toolname drops ZERO events while a super-producer is running. In Figure~\ref{fig:droppaper}, the line for \toolname overlaps with the diagonal line, indicating that no events have been dropped. In contrast, existing pro-performance solutions drop most of the generated system events, allowing for a high success rate for \ac{pdos} attacks. Specifically, on machines with 1 CPU core and 2 GB memory, Sysdig drops 31\% and 33\% of system call events on virtual and physical machines, respectively. On 32-core machines, Sysdig drops 98.5\% and 98.9\% of system call events on virtual and physical machines, respectively. Similarly, LTTng drops 71.9\% of total system call events on a 32-core virtual machine and about 60\% of total events on a 32-core physical machine. LTTng does not drop events on machines with 1 CPU core and 2 GB memory because it adopts a dynamic buffer mechanism that can hold more events than Sysdig. However, as we will show in the next section, this design also causes LTTng to introduce more system overhead than Sysdig. For Linux Audit, since it is not as well optimized as Sysdig and LTTng~\cite{lttng}, it drops nearly all system call events when the super producer reaches 1\% of its highest event generation speed.

\subsubsection{Preventing \ac{pdos} in realistic web-apps:} 
We evaluated whether \toolname can prevent \ac{pdos} attacks by simulating a production-level web-server. We used the 'web-serving' benchmark from CloudSuite as the implementation of the \textit{victim app}. This benchmark hosts a production-quality social networking engine, which we hosted with Apache 2.4, MariaDB 10.1, PHP 7.4, and Elgg 3.3. The \textit{target app} was implemented as a static website that provides a "ping lookup" service to users, but it had a command line injection vulnerability that allowed attackers to run remote code. We hosted the \textit{target app} in a different process of Apache and evaluated \ac{pdos} in two widely adopted resource isolation methods: (1) the \textit{target app} and \textit{victim app} were scheduled and isolated by the default Linux configuration, which focuses on maximizing system utilization while providing fairness and performance isolation in a best-effort way; and (2) the \textit{target app} and \textit{victim app} were in different \code{cgroups} with isolated CPU utilization. All applications were deployed on an Intel Xeon Silver server with 16 CPU cores, 128 GB memory, and a 10 TB HDD.

Our implemented \ac{pdos} attack consists of three steps. First, we simulate 20 visitors accessing the \textit{victim app} and turn it into a super producer. Second, the attacker waits for three to five minutes to ensure that the auditing frameworks are overloaded. Third, the attacker initiates a command \& control connection to the \textit{target app} by exploiting a command line injection vulnerability. We consider the \ac{pdos} attack successful if the auditing frameworks do not record any provenance data about the command \& control connection. For each auditing framework, we conduct 120 \ac{pdos} attacks and count how many of them are successful.

\noindent\textbf{Results:} Our experiment shows that \toolname can prevent \ac{pdos} attacks, while all three baseline methods are vulnerable to the \ac{pdos} attack. The attack success rates for these baseline methods are higher than 90\% in all cases, with at least 107 out of 120 \ac{pdos} attacks being successful. These results are shown in Table~\ref{tab:pdosattack} for both isolation methods. Furthermore, using \code{cgroup} does not prevent the \ac{pdos} attack because the \textit{victim app} does not exceed its quota, and the collector is unable to isolate provenance events internally.
 
\begin{table}[t]
    \caption{\small Attack success rate of the \ac{pdos} attack ({\#}successful/ {\#}attempts)}
     \vspace{-0.5em}    
    \label{tab:pdosattack}
    \centering
       \resizebox{0.49\textwidth}{!}{
    \begin{tabular}{|c|c|c|c|c|}
        \hline
        & Sysdig &LTTng &Linux Audit & \cellcolor[gray]{0.8}\toolname\\ \hline
    Default    & 120/120    & 107/120   &  120/120 & \cellcolor[gray]{0.8}0/120 \\\hline
    Cgroup     & 115/120    & 107/120   &  120/120 & \cellcolor[gray]{0.8} 0/120\\
    \hline
    \end{tabular}
    }
    \vspace{-1em}    
\end{table}

\begin{figure}
\centering
\includegraphics[scale=1]{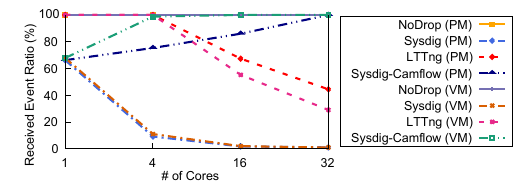}

\vspace{-0.1in}
\caption{Summary of how the auditing frameworks drop events with different hardware configurations. \emph{The line of \toolname overlaps with the 100\% events received line, indicating no event has been dropped across all configurations.}}
\label{fig:summarypaper}
\vspace{-1.5em}
\end{figure}

\subsection{RQ 2: Application Slowdown}
\label{ssec:slowdonw}
To answer this research question, we implemented realistic \ac{pados} attacks using realistic web applications. In the \ac{pados} attack, we assumed that the \textit{target app} and the \textit{victim app} were running in different \code{cgroups}. We implemented the \textit{target app} and the \textit{victim app} in different \code{cgroups} to allow for absolute isolation between them. An auditing framework was running in both \code{cgroups} to monitor the \textit{target app} and the \textit{victim app}. 

We implemented the \textit{victim app} using the same script as in \S\ref{ssec:eventdrop}, and we implemented the \textit{target apps} using three widely used web applications: Nginx~\cite{Nginx}, Redis~\cite{redis}, and OpenSSL~\cite{openssl}. By leveraging the script, we were able to control the generation speed of provenance data in the super producer (the \textit{victim app}). We evaluated Sysdig, Linux Audit, and LTTng to show how they affected the performance of the \textit{target app}. To adopt a pro-integrity strategy, we implemented two additional versions based on Sysdig. The first one, called Sysdig-Camflow, optimized Camflow using a more efficient kernel module from Sysdig while preserving the per-core thread user-space collector from Camflow. The second one, called Sysdig-Integrity, integrated synchronized event processing~\cite{ahmad2022hardlog} into Sysdig to ensure the integrity of provenance data. Sysdig-Integrity blocked the currently running process when its event buffer was full and woke it up once the buffer had been processed. In this way, Sysdig-Integrity ensured zero event loss and was a guaranteed pro-integrity solution.

We also reported the performance of applications on a vanilla machine as the "No Consumer" to show the baseline performance of the system without any auditing frameworks in the user space. To measure the system performance of our applications, we ran their corresponding benchmark scripts and reported the official performance scores reported by the scripts. This avoided statistical data bias that could have resulted from poorly self-implemented benchmark scripts.

\begin{figure*}[!t]
    \centering
    \includegraphics[width=\textwidth]{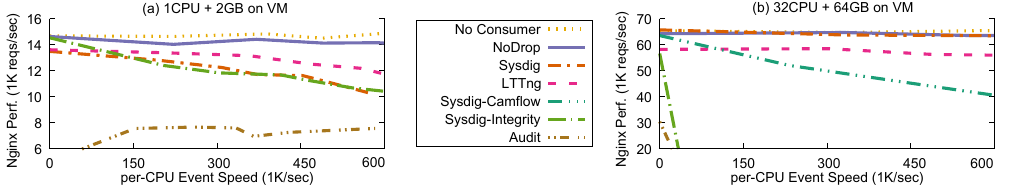}
    \vspace{-0.25in}
    \caption{The performance of Nginx on different platforms and hardware configurations under various workloads from the super producer. \emph{The x-axis is the speed of generated events in 30 seconds, and the y-axis is the performance of Nginx, measured in the number of requests per second.}}
    \label{fig:rq12-nginx-1+32}
    \vspace{-1em}
\end{figure*}

Figure~\ref{fig:rq12-nginx-1+32} shows detailed results for Nginx with 1 CPU core and 32 CPU cores on virtual machines. The results for other configurations are similar, but we have included them in our GitHub repository due to space constraints. The x-axis of Figure~\ref{fig:rq12-nginx-1+32} represents the workload of the super producer, namely the speed of generated events per second, while the y-axis represents the official performance score reported by the benchmark scripts. Since the number of consumer threads in Sysdig-Camflow equals the number of CPU cores, Sysdig-Camflow and Sysdig are identical on a single-core machine. Therefore, we merged the lines for Sysdig-Camflow and Sysdig in Figure~\ref{fig:rq12-nginx-1+32}(a). Overall, our experiment showed that an attacker could paralyze the \textit{target app} by turning the \textit{victim} into a super producer in a different \code{cgroup}.

\begin{figure*}[!t]
    \centering
    \includegraphics[width=\textwidth]{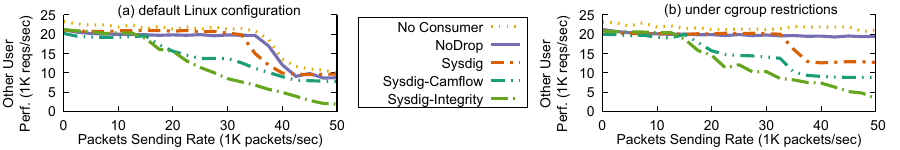}
    \vspace{-0.26in}
    \caption{\small The performance of the \textit{target app}(static page) in the case CloudSuite with a different workload from the super producer.\emph{The x-axis is the packet sending rate of the super producer, and the y-axis is the performance score of \textit{wrk}, measured in the number of returned HTTP responses. Higher is better.}}
    \label{fig:padoscase}
    \vspace{-0.5em}
\end{figure*}

Our experiments showed that (1) \toolname prevents the \ac{pados} attack, (2) existing pro-integrity solutions suffer from the \ac{pados} attack across different hardware configurations, and (3) pro-performance collectors also slow down the \textit{target app}. For all three web applications on all eight hardware configurations, when the workload of the super producer increased, \toolname maintained stable application performance regardless of the increasing workload from the super producers. For example, in Figure~\ref{fig:rq12-nginx-1+32}, \toolname was at most 5.1\% slower on a single-core virtual machine and at most 0.5\% slower on a 32 CPU cores virtual machine than the ideal "No Consumer" baseline. This overhead was also lower than that of the baselines. This result proves that \toolname is robust against \ac{pados} attacks. In other words, an attacker cannot slow down applications in different \code{cgroup} by running a super producer.

We also notice that the pro-performance collectors can also be vulnerable to the \ac{pados} attack. On the single-core virtual and physical machines, Sysdig, LTTng, and Linux Audit also decrease the performance of the \textit{target app} proportionally to the event generation speed of the super producer. This is because Sysdig, LTTng, and Linux Audit limit the resource usage of their user-space component by allowing only one user-space thread. Thus, Sysdig, LTTng, and Linux Audit behave the same as Sysdig-Camflow on a single-core platform, making them de facto pro-integrity collectors. Linux Audit has the worst performance because it relies on Netlink~\cite{salim2003linux} to pass events from the kernel to the user-space component, which leads to a less efficient kernel module than other auditing frameworks.

\subsubsection{CloudSuite Setting} 
To further evaluate how well \toolname can prevent \ac{pados} attacks in a production environment, we also implemented \ac{pados} attacks using the CloudSuite setting that we used in \S\ref{ssec:eventdrop}.

Our implemented \ac{pados} attack consisted of three steps. First, similar to \ac{pdos}, the attacker used 20 visitors to generate a flood of remote requests to the victim app. We further adjusted the workload of visitors to generate increasing pressure on the victim app. Second, the attacker waited for five minutes to ensure that the auditing framework had enough resources. Third, the attacker started a normal DoS attack on the target app. In this case, the required workload for the DoS attack on the target app was substantially reduced.

In our experiment, we measured the performance reduction of the target app to evaluate the effectiveness of the \ac{pados} attack. To accurately measure the performance of the target app, we also used \textit{wrk}~\cite{wrk} to issue HTTP requests to the target app and reported the number of returned HTTP responses as its performance metric.

\textbf{Results:} Our evaluation showed that pro-integrity collectors could substantially decrease the performance of the \textit{target app} and, thus, amplify possible DoS attacks. We showed the performance of the \textit{target app} in Figure~\ref{fig:padoscase}. Note that the request sending rate on the x-axis represents the speed at which requests reach the \textit{victim app}, not the request processing speed of the app. To better understand how pro-integrity collectors decrease the performance of the \textit{target app}, we plotted the case of Sysdig and the case of no provenance collectors (No Consumer) in Figure~\ref{fig:padoscase} as baselines.

With the default Linux configuration (Figure~\ref{fig:padoscase}(a)), we observed that pro-integrity collectors accelerated the DoS attack, requiring less workload to slow down the \textit{target app}. Specifically, the curve for No Consumer remained flat when the system had sufficient spare resources to digest attack traffic. When the workload exceeded 34K packets/s, however, the whole system became overloaded, and the performance of the \textit{target app} rapidly decreased. However, when provenance collectors were present, the \textit{target app} had smaller turning points (for Sysdig, Sysdig-Camflow, and Sysdig-Integrity, these turning points were 32K, 15K, and 15K, respectively) and much lower performance than No Consumer. This was because pro-integrity collectors consumed too many system resources and competed with other applications, overloading the server more easily. In general, Figure~\ref{fig:padoscase}(a) shows that pro-integrity collectors can amplify DoS attacks since they blindly compete for resources with other applications.
 
Figure~\ref{fig:padoscase}(b) reports the results under \code{cgroup} restrictions. The \textit{target app} and the \textit{victim app} had 20\% and 80\% CPU utilization limitations, respectively. Consequently, the super producer could not use up all system resources, and the \textit{target app} was not impacted. The flat line for No Consumer in Figure~\ref{fig:padoscase}(b) indicates this, showing that the DoS attack was ineffective. However, pro-integrity collectors compromised the isolation of \code{cgroup} and greatly influenced the \textit{target app}. In the case of Sysdig, Sysdig-Camflow, and Sysdig-Integrity, the worst performance loss for the \textit{target app} reached 40.1\%, 55.6\%, and 83.7\%, respectively. Please note that placing a \code{cgroup} limitation on provenance collectors would cause a \ac{pados} attack as studied above. In contrast to existing provenance collectors, \toolname preserved \code{cgroup} isolation and successfully protected the \textit{target app} from DoS attacks. 

\subsection{RQ 3: Runtime Overhead}
\label{sec:overhead}
This research question evaluated the runtime overhead of \toolname by measuring its impact on the OS and on running applications, both with and without a super producer. We used Sysdig as a baseline, repeated measurements ten times, and employed statistical methods, including the Wilcoxon signed-rank test, to test for significant differences between Sysdig and \toolname. We used p-value~\cite{Pvalue} as our evaluation metric and concluded that results were statistically significant when $p \textless 0.05$.

\subsubsection{OS Overhead}
\label{ssec:syscallovehead}
Similar to evaluations in other auditing frameworks~\cite{paccagnella2020logging}, we used \textit{lmbench}~\cite{mcvoy1996lmbench} to measure the OS overhead of \toolname. On average, the OS overhead of \toolname was 35\% higher than vanilla Linux across eight different hardware configurations. We also showed the relative overhead percentage compared to Sysdig for four configurations in Table~\ref{tab:lmbench} and left the results for other configurations in our GitHub repository.

Our experiments show that the OS overhead of \toolname is similar to that of Sysdig. On average, the OS overhead of \toolname is 0.2\% lower than Sysdig across eight different hardware configurations. Overall, we cannot find a statistically significant difference between Sysdig and \toolname. We conclude that the differences between the OS overhead of \toolname and Sysdig are mainly due to the randomness of a dynamic system. The OS overhead measures the performance of the kernel modules of \toolname and Sysdig. Since \toolname implements its kernel module in the same way as Sysdig, their OS overhead should be the same.

\begin{table}[h]
    \centering
    \footnotesize
    \caption{\small Performance scores of \textit{lmbench}. \emph{All values are shown as percentages relative to Sysdig. The negative value means \toolname is faster than Sysdig.}}
      \resizebox{0.5\textwidth}{!}{
    \begin{tabular}{lrrrrr}
    \toprule
Configurations & 
    C1 & C4 & C5 & C8 & Ave  \\ 
    \cmidrule{1-6}
    \multicolumn{6}{c}{\small Syscall Tests} \\
    \cmidrule{0-1}\cmidrule{2-6}
NULL syscall & 
    -8.1\% & -17\% & -8.3\% & -7.9\% & -10.3\% \\ 
stat & 
    -9.0\% & +5.5\% & -1.8\% & -0.6\% & -1.5\% \\
fstat & 
    +4.2\% & -1.7\% & +1.7\% & +1.6\% & +2.3\% \\ 
open/close file & 
    -6.1\% & -2.9\% & -0.3\% & -1.8\% & -2.8\% \\ 
read file & 
    +7.4\% & +7.1\% & +4.5\% & +7.2\% & +6.6\% \\ 
write file & 
    +7.7\% & +7.2\% & +12.5\% & +12.1\% & +9.9\% \\ 
    \cmidrule{0-1}\cmidrule{2-6}
    \multicolumn{6}{c}{\small File Access} \\
    \cmidrule{0-1}\cmidrule{2-6}
file create (0K) & 
    -15.8\% & -7.1\% & -10.0\% & +2.7\% & -7.5\% \\ 
file delete (0K) & 
    +0.5\% & +3.0\% & -0.7\% & -0.9\% & +0.5\% \\ 
file create (10K) & 
    +0.1\% & +2.9\% & -3.7\% & -0.8\% & -0.4\% \\ 
file delete (10K) & 
    +4.7\% & +1.5\% & -0.9\% & -0.4\% & +1.2\% \\ 
pipe & 
    +3.0\% & +0.8\% & +6.9\% & +1.3\% & +3.0\% \\ 
AF\_UNIX & 
    +3.8\% & -10.5\% & +5.3\% & +10.1\% & +2.2\% \\ 
    \bottomrule
    \end{tabular}
    }
\label{tab:lmbench}
\vspace{-1em}
\end{table}

\subsubsection{Application Overhead}
\label{sec:appoverhead}
To answer this research question, we conducted macro-benchmarks to measure the performance of seven applications, both with and without a super producer. These applications can be divided into two categories: the first category includes I/O-intensive benchmarks such as Nginx\cite{Nginx}, Redis\cite{redis}, and Postmark\cite{katcher1997postmark}, as well as two other applications, Django~\cite{py-django} for Python and http~\cite{go-http} for Golang. The second category includes CPU-intensive benchmarks, namely OpenSSL~\cite{openssl} and 7-ZIP~\cite{7zip}.

\begin{table*}[!ht]
\centering
\label{tab:new}
\begin{subtable}{0.45\textwidth}
\centering
\caption{\textbf{Benchmark WITHOUT super producer.}}
\label{tab:benchmarkidle}
    \begin{tabular}{l|l|rrrr} 

\hline
\textbf{Application}             & \textbf{Collector}                       & \textbf{C1}                                & \textbf{C4}                               & \textbf{C5}                                & \textbf{C8}                                               \\ 
\hline\hline
\multirow{3}{*}{Nginx}           & \toolname                                   & 9.80                                                & 3.60                                               & 11.35                                               & 4.84                                                \\
                                 & Sysdig                                   & 55.30                                               & 5.50                                               & 37.64                                               & 7.20                                                \\ 
\cline{2-6}
                                 & {\cellcolor[rgb]{0.906,0.906,0.906}}DIFF & {\cellcolor[rgb]{0.906,0.906,0.906}}\textbf{-29.30} & {\cellcolor[rgb]{0.906,0.906,0.906}}\textbf{-1.80} & {\cellcolor[rgb]{0.906,0.906,0.906}}\textbf{-19.10} & {\cellcolor[rgb]{0.906,0.906,0.906}}\textbf{-2.20}  \\ 
\hline
\multirow{3}{*}{Redis}           & \toolname                                   & 8.90                                                & 3.10                                               & 8.66                                                & 2.42                                                \\
                                 & Sysdig                                   & 21.00                                               & 5.00                                               & 21.00                                               & 5.70                                                \\ 
\cline{2-6}
                                 & {\cellcolor[rgb]{0.906,0.906,0.906}}DIFF & \textbf{{\cellcolor[rgb]{0.906,0.906,0.906}}-11.10 }         & \textbf{ {\cellcolor[rgb]{0.906,0.906,0.906}}-2.00  }        & \textbf{ {\cellcolor[rgb]{0.906,0.906,0.906}}-10.20    }      & \textbf{{\cellcolor[rgb]{0.906,0.906,0.906}}-3.10      }   \\ 
\hline
\multirow{3}{*}{Postmark}        & \toolname                                   & 25.50                                               & 19.10                                              & 30.60                                               & 20.65                                               \\
                                 & Sysdig                                   & 96.30                                               & 8.60                                               & 95.80                                               & 13.50                                                \\ 
\cline{2-6}
                                 & {\cellcolor[rgb]{0.906,0.906,0.906}}DIFF & {\cellcolor[rgb]{0.906,0.906,0.906}}\textbf{-36.00} & {\cellcolor[rgb]{0.906,0.906,0.906}}\textbf{9.70}  & {\cellcolor[rgb]{0.906,0.906,0.906}}\textbf{-33.30} & {\cellcolor[rgb]{0.906,0.906,0.906}}\textbf{6.30}   \\ 
\hline
\multirow{3}{*}{\makecell{Django \\ (Python)}} & \toolname                                   & 1.30                                                & 2.00                                               & 1.30                                                & -0.20                                               \\
                                 & Sysdig                                   & 1.10                                                & 2.30                                               & 1.10                                                & 0.30                                                \\ 
\cline{2-6}
                                 & {\cellcolor[rgb]{0.906,0.906,0.906}}DIFF & {\cellcolor[rgb]{0.906,0.906,0.906}}0.30            & {\cellcolor[rgb]{0.906,0.906,0.906}}-0.30          & {\cellcolor[rgb]{0.906,0.906,0.906}}0.20            & {\cellcolor[rgb]{0.906,0.906,0.906}}-0.50           \\ 
\hline
\multirow{3}{*}{\makecell{http \\ (Golang)}}   & \toolname                                   & 14.10                                               & 2.20                                               & 14.66                                               & 2.71                                                \\
                                 & Sysdig                                   & 78.90                                               & 2.20                                               & 65.70                                               & 2.20                                                \\ 
\cline{2-6}
                                 & {\cellcolor[rgb]{0.906,0.906,0.906}}DIFF & {\cellcolor[rgb]{0.906,0.906,0.906}}\textbf{-36.20} & {\cellcolor[rgb]{0.906,0.906,0.906}}0.10           & {\cellcolor[rgb]{0.906,0.906,0.906}}\textbf{-30.80} & {\cellcolor[rgb]{0.906,0.906,0.906}}0.50            \\ 
\hline
\multirow{3}{*}{OpenSSL}         & \toolname                                   & 0.60                                                & 0.10                                               & 0.20                                                & 0.70                                                \\
                                 & Sysdig                                   & 0.60                                                & 0.10                                               & 0.20                                                & 0.60                                                \\ 
\cline{2-6}
                                 & {\cellcolor[rgb]{0.906,0.906,0.906}}DIFF & {\cellcolor[rgb]{0.906,0.906,0.906}}0.10            & {\cellcolor[rgb]{0.906,0.906,0.906}}0.00           & {\cellcolor[rgb]{0.906,0.906,0.906}}0.00            & {\cellcolor[rgb]{0.906,0.906,0.906}}0.10            \\ 
\hline
\multirow{3}{*}{7-ZIP}           & \toolname                                   & 0.30                                                & 0.80                                               & 1.20                                                & 0.70                                                \\
                                 & Sysdig                                   & 0.20                                                & 0.70                                               & 1.20                                                & 0.70                                                \\ 
\cline{2-6}
                                 & {\cellcolor[rgb]{0.906,0.906,0.906}}DIFF & {\cellcolor[rgb]{0.906,0.906,0.906}}0.10            & {\cellcolor[rgb]{0.906,0.906,0.906}}0.10           & {\cellcolor[rgb]{0.906,0.906,0.906}}0.00            & {\cellcolor[rgb]{0.906,0.906,0.906}}0.00            \\ 
\hline
\multirow{3}{*}{PostgreSQL}      & \toolname                                   & 7.05                                                & 3.80                                               & 10.20                                               & 4.70                                                \\
                                 & Sysdig                                   & 15.20                                               & 4.70                                               & 17.40                                               & 4.90                                                \\ 
\cline{2-6}
                                 & {\cellcolor[rgb]{0.906,0.906,0.906}}DIFF & \textbf{{\cellcolor[rgb]{0.906,0.906,0.906}}-7.61 }          & {\cellcolor[rgb]{0.906,0.906,0.906}}-0.80           & \textbf{{\cellcolor[rgb]{0.906,0.906,0.906}}-6.50   }        & {\cellcolor[rgb]{0.906,0.906,0.906}}-0.19           \\
\hline
\end{tabular}
\end{subtable}
\hspace{1cm}
\begin{subtable}{0.45\textwidth}
\centering
\caption{\textbf{Benchmark WITH super producer.}}
\label{tab:benchmarkbusy}
\begin{tabular}{rrrr} 
\hline
 \textbf{C1}                                & \textbf{C4}                               & \textbf{C5}                                & \textbf{C8}                                \\ 
\hline\hline
                               10.00                                      & 3.20                                      & 13.15                                      & 1.96                                       \\
                                                         51.93                                      & 6.28                                      & 58.70                                      & 3.20                                       \\ 
\cline{1-4}
                                 \textbf{{\cellcolor[rgb]{0.906,0.906,0.906}}-27.60 }& \textbf{{\cellcolor[rgb]{0.906,0.906,0.906}}-2.90} & \textbf{{\cellcolor[rgb]{0.906,0.906,0.906}}-28.70 }& \textbf{{\cellcolor[rgb]{0.906,0.906,0.906}}-1.20  }\\ 
\hline
                               3.70                                       & 1.20                                      & 4.23                                       & 0.27                                       \\
                                                  56.17                                      & 7.66                                      & 61.10                                      & 3.80                                       \\ 
\cline{1-4}
                          \textbf{{\cellcolor[rgb]{0.906,0.906,0.906}}-33.60} & \textbf{{\cellcolor[rgb]{0.906,0.906,0.906}}-6.00 }& \textbf{{\cellcolor[rgb]{0.906,0.906,0.906}}-35.30 }& \textbf{{\cellcolor[rgb]{0.906,0.906,0.906}}-3.40  }\\ 
\hline
                                   15.10                                      & 33.50                                     & 14.30                                      & 30.20                                      \\
                                                              65.37                                      & 6.37                                      & 68.58                                      & 7.43                                       \\ 
\cline{1-4}
                                 \textbf{{\cellcolor[rgb]{0.906,0.906,0.906}}-30.40} & \textbf{{\cellcolor[rgb]{0.906,0.906,0.906}}25.50 }& \textbf{{\cellcolor[rgb]{0.906,0.906,0.906}}-32.20} & \textbf{{\cellcolor[rgb]{0.906,0.906,0.906}}21.20  }\\ 
\hline
                               1.20                                       & 3.30                                      & 1.50                                       & 1.40                                       \\
                                                             41.14                                      & 2.18                                      & 47.74                                      & 0.90                                       \\ 
\cline{1-4}
                             \textbf{{\cellcolor[rgb]{0.906,0.906,0.906}}-28.30} & {\cellcolor[rgb]{0.906,0.906,0.906}}1.10  & \textbf{{\cellcolor[rgb]{0.906,0.906,0.906}}-31.30} & {\cellcolor[rgb]{0.906,0.906,0.906}}0.50   \\ 
\hline
                              5.80                                       & 4.80                                      & 9.80                                       & 3.20                                       \\
                                                        47.56                                      & 0.48                                      & 47.38                                      & 0.10                                       \\ 
\cline{1-4}
                         \textbf{{\cellcolor[rgb]{0.906,0.906,0.906}}-28.30} & \textbf{{\cellcolor[rgb]{0.906,0.906,0.906}}4.30}  & \textbf{{\cellcolor[rgb]{0.906,0.906,0.906}}-25.50} & \textbf{{\cellcolor[rgb]{0.906,0.906,0.906}}3.10  } \\ 
\hline
                                0.80                                       & 3.20                                      & 0.17                                       & 0.08                                       \\
                                            47.37                                      & 5.09                                      & 43.30                                      & 1.40                                       \\ 
\cline{1-4}
                             \textbf{{\cellcolor[rgb]{0.906,0.906,0.906}}-31.60} & \textbf{{\cellcolor[rgb]{0.906,0.906,0.906}}-1.80} & \textbf{{\cellcolor[rgb]{0.906,0.906,0.906}}-30.10} & \textbf{{\cellcolor[rgb]{0.906,0.906,0.906}}-1.30 } \\ 
\hline
                                   0.40                                       & 1.50                                      & 0.20                                       & 1.20                                       \\
                                                      50.30                                      & 4.86                                      & 38.02                                      & 3.16                                       \\ 
\cline{1-4}
                              \textbf{{\cellcolor[rgb]{0.906,0.906,0.906}}-33.20} & \textbf{{\cellcolor[rgb]{0.906,0.906,0.906}}-3.20 }& \textbf{{\cellcolor[rgb]{0.906,0.906,0.906}}-27.40} & \textbf{{\cellcolor[rgb]{0.906,0.906,0.906}}-1.90 } \\ 
\hline
                                 11.60                                      & 4.50                                      & 12.20                                      & 6.20                                       \\
                                                               22.30                                      & 4.86                                      & 25.02                                      & 6.30                                       \\ 
\cline{1-4}
                \textbf{{\cellcolor[rgb]{0.906,0.906,0.906}}-9.50 } & {\cellcolor[rgb]{0.906,0.906,0.906}}-0.34 & \textbf{{\cellcolor[rgb]{0.906,0.906,0.906}}-11.40} & {\cellcolor[rgb]{0.906,0.906,0.906}}0.00   \\
\hline
\end{tabular}
\end{subtable}
\caption{\textbf{We measured the processing time per request/transaction for seven representative applications and a Kubernetes-based PostgreSQL. For each application, the first two lines show the relative runtime overhead (\%) compared to vanilla Linux, where a lower value indicates performance closer to that of vanilla Linux. The third line shows the relative overhead between \toolname and Sysdig, with values smaller than 0 indicating that \toolname outperforms Sysdig. For brevity, we denote this as DIFF. We report the mean values across 10 runs, with p-values less than 0.05 shown in bold.}}
\end{table*}

We use the official benchmark tools with their default settings for different hardware configurations to evaluate their performance. We repeat each experiment 10 times and measure the average metrics reported by the benchmarks of each application. Specifically, We use \textit{wrk}~\cite{wrk} configured with 1,000 concurrent connections to benchmark Nginx. For Redis, we use the \textit{redis-benchmark} configured to send 1,000,000 requests and measure the speed of operation \code{get}. For Postmark, we use the built-in benchmark with the configuration of manipulating 500 files concurrently and launching 100,000 transactions. We rely on the Phoronix Test Suite, one of the most comprehensive benchmark suites of web applications~\cite{phoronix,sharath-ijca-13}, to benchmark Django and http. For OpenSSL, We use the built-in \textit{speed} benchmark configured to utilize all CPU cores and measure the time to compute one rsa4096 signature. For 7-ZIP, we use the built-in benchmark configured to utilize all CPU cores and measure the compression speed in MIPS. We report the relative cost of \toolname to Sysdig for four configurations in Table~\ref{tab:benchmarkidle} (without the super producer) and Table~\ref{tab:benchmarkbusy} (with the super producer). We also measure the overhead relative to the vanilla Linux with no auditing framework running at all. We leave the results of other configurations in our GitHub repository.

On average, the overhead of \toolname is 6.58\% higher than vanilla Linux and 6.30\% lower than Sysdig across eight different configurations. Our statistical analysis confirms the validity of our data. Both Sysdig and \toolname introduce overhead compared to vanilla Linux since they record and consume provenance events. However, the overheads of \toolname to vanilla Linux are relatively less, with overheads less than 15\% for all applications except Postmark. The reasons for the difference in application overheads between \toolname and Sysdig depend on hardware configurations and application categories.

For single-core machines (C1 and C5 in Table~\ref{tab:benchmarkidle} and Table~\ref{tab:benchmarkbusy}), \toolname is more efficient because it eliminates the process scheduling overhead. With a single-core, the OS needs to periodically switch to the Sysdig process for provenance data processing, which leads to higher application overhead. 

For multi-core machines, \toolname shows relatively high performance because it not only eliminates the process switching cost of Sysdig but also avoids cross-core data transmission. With multiple cores, the kernel module collects the provenance data on the same core as the running application. However, the Sysdig process accesses the data in parallel on a different core. This introduces a notable overhead of cache coherence across the two cores due to the shared provenance data buffer. On the contrary, \toolname processes the provenance data on the same core as the in-kernel collector, avoiding the cost of cross-core data transmission.

As shown in Tables~\ref{tab:benchmarkidle} and~\ref{tab:benchmarkbusy}, \toolname offers lower runtime overhead than Sysdig for applications with many I/O-intensive processes, such as Nginx, Redis, and http. If there are more I/O-intensive processes than CPU cores, the Sysdig auditing processes will compete for computational resources with the monitored applications. In other words, the auditing processes of Sysdig will interrupt the monitored apps in the same manner as \toolname. However, \toolname employs threadlets by design, which add less scheduling overhead than Sysdig. The fewer the number of CPU cores, the greater the scheduling overhead for Sysdig. Thus, \toolname is typically more effective when there are fewer CPU cores. Moreover, we find that as the number of CPU cores increases, the application's event generation speed decreases due to changes in application architecture. As a result, \toolname offers lower runtime overhead in C4 compared to C1. 

For CPU-intensive applications such as 7-ZIP, OpenSSL, and Django, they generate almost no system call events. Therefore, when there is no super producer, the overheads of both \toolname and Sysdig compared to vanilla Linux are much smaller, averaging less than 2\% across all configurations. When a super producer is running, the overhead of \toolname remains while the overhead of Sysdig increases. This is because the centralized Sysdig auditing process needs to process a large number of events generated by the super producer and will persistently compete for computational resources with the monitored application.

Sysdig may introduce less runtime overhead to the monitored process when there are spare CPU cores available to host the centralized auditing processes. In this case, there is no resource competition between the monitored and auditing processes. This is also the case for Postmark in C4 of Tables~\ref{tab:benchmarkidle} and~\ref{tab:benchmarkbusy} since our Postmark benchmark is single-threaded. Although \toolname shows higher runtime overhead for Postmark compared to Sysdig, it spares the core that would otherwise be used to host the Sysdig auditing process and prevents event dropping.

\noindent\textbf{Kubernetes-managed application.} To monitor IO-bound applications and complex systems, we combine PostgreSQL Operator with Pgpool-II to deploy a PostgreSQL cluster with query load balancing and connection pooling capability on Kubernetes ~\cite{pgpool}. We deploy a Pgpool-II pod that contains a Pgpool-II container and a Pgpool-II Exporter container. The Pgpool-II container Docker image is built with streaming replication mode. We set the replicas to 1, so we have three pods in total. To test the replication functionality, we use a benchmark tool called pgbench ~\cite{pgbench}, which comes with the standard PostgreSQL installation, and we measure the processing time per transaction. We repeat each test 10 times for all configurations, and each test lasts for 20 seconds.

Since we implemented the privilege escalation as mentioned in \S\ref{sec:consumer}, \toolname is able to monitor system behaviors inside the Docker container. The results are shown in Tables~\ref{tab:benchmarkidle} and~\ref{tab:benchmarkbusy}.  We find that for each configuration, \toolname introduces less than 15\% overhead compared to vanilla Linux. On average, \toolname introduces 3.34\% lower overhead than Sysdig. These results show that \toolname is suitable for monitoring Kubernetes-based deployments.

\subsection{RQ 4: Effectiveness of Data Reduction}
Several log reduction and partitioning techniques, such as CPR~\cite{10.1145/2976749.2978378}, LogGC~\cite{10.1145/2508859.2516731}, ProTracer~\cite{MaZX16}, and KCAL~\cite{10.5555/3277355.3277379}, do not solve the data integrity vs. performance dilemma because they add high computation overhead, amplifying PADoS attacks. To validate their ineffectiveness, we modified Sysdig's code by inlining the CPR algorithm into the kernel and called it Sysdig-CPR.

\textbf{Design of Sysdig-CPR:} Since the CPR algorithm is an offline algorithm that depends on the global properties of graphs, we maintain a temporary graph in an extra 8M kernel buffer for each CPU core. The design of Sysdig-CPR is similar to Sysdig-Integrity, which blocks the currently running process when the event buffer (kernel and userspace shared buffer) is full and wakes it up once the buffer has been processed. Sysdig-CPR does two additional things: it wakes up a kernel thread to run the CPR algorithm when the kernel buffer is full and copies the reduced events to the shared buffer for consumption; and it notifies the userspace component to consume the buffer. Sysdig-CPR is available at: \url{https://github.com/nodropforsecurity/sysdigcpr}

\textbf{Results:} Our measurement follows the configurations of \S\ref{ssec:slowdonw}. We omit Sysdig-CPR for clarity in Figure~\ref{fig:rq12-nginx-1+32} as well as our GitHub repository because we find that the performance curves of Sysdig-CPR in all configurations are stuck on the horizontal axis. In our experiment, the kernel CPR can handle 2,000 events per second per core, which is consistent with the original paper~\cite{10.1145/2976749.2978378}. Although it can reduce most of the events (more than 70\%), the super producer can easily generate 100,000 events per second. This means that the system will take several seconds to handle the generated events, which greatly blocks the running applications.

\subsection{RQ 5: Effectiveness of Increasing Buffer Size}

One possible approach to avoid event dropping in system auditing frameworks~\cite{sysdigcve,discuss1,discuss2} is to increase the buffer size. However, this approach is ineffective. We applied this approach to Sysdig, LTTng, and Linux Audit and evaluated them as follows. For each CPU core, we set the maximum applicable buffer size, which is 768M for Sysdig and LTTng and 77,000 messages for Linux Audit. We cannot increase the buffer size even larger because the system will crash when the buffer size exceeds the threshold. In our measurement shown in Table~\ref{tab:collectorbuffer}, Sysdig, LTTng, and Linux Audit still drop events with the same super producer configuration in \S\ref{ssec:eventdrop}. The dropping rate is 88\%, 52\%, and 99\% for Sysdig, LTTng, and Linux Audit when the generation speed per core reaches 1.6 million per second. Moreover, as reported by the developers of Sysdig, increasing the size of the ring buffer may significantly slow down the whole system~\cite{sysdigcve,buffersizeslow}. According to our experiment, when increasing buffer size from 8M to 768M, the events generation speed decreases by 35\% under the same stress test.

\begin{table}[t]
\setlength{\abovecaptionskip}{5pt}
 \caption{Dropping rate of auditing frameworks with maximum buffer size.}
    \label{tab:collectorbuffer}
\begin{tabular}{|l|c|c|c|c|}
\hline
\textbf{Configuration}  & \textbf{C1}   & \textbf{C4}   & \textbf{C5}  &\textbf{C8}  \\ \hline
Linux Audit &  99.2\% & 99.6\% & 99.1\%& 99.4\%\\ \hline
Sysdig  &  19.3\% & 86.1\% & 25.5\% & 88.1\%\\ \hline
LTTng  &  0\% & 49.5\% & 0\% & 52.1\%\\ \hline
\end{tabular}
\end{table}

\section{Related Work}

Provenance analysis has been widely applied in different security tasks, such as APT attack investigation~\cite{LPM,Spade,king_backtracking_2003-1,MaZX16,camflow,hifi,gao2018saql,gao2018aiql,ji2017rain,ji2018enabling,liu2018towards,pasquier2018ccs} and detecting stealthy security risks~\cite{berlin2015malicious,du2017deeplog,han2020ndss,gu2015leaps,hassan2019nodoze,hossain_sleuth_2017,liu2019log2vec,milajerdi2019holmes,milajerdi2019poirot,oprea2015detection,pei2016hercule,shen2018tiresias,shen2019attack2vec,wang2020you,9152771,han2020sigl}.  There are also methods for precisely and clearly interpreting events to explain applications' behaviors~\cite{bates2017transparent,hassan2020omegalog,kwon2018mci,beep,ma2015accurate,ma2017mpi,Pass,yang2020uiscope}. \toolname benefits these tasks by providing a more reliable data source. Attackers regularly engage in anti-forensic activities to cover their tracks~\cite{asi}. Several cryptographic-based approaches are proposed to secure logs~\cite{paccagnella2020logging,Forwardsecure,syslog-ng,10.1145/3052973.3053034,Custos}, but none discuss the security of the user-space component of auditing frameworks.

Threadlet is a short sequence of instructions with self-contained memory~\cite{kogge2004piglets,4228404,9188165,9622823,9652820}. \toolname borrows this concept but implements it differently as a piece of code instrumented to a host thread. \toolname provides protections such as MPK, address randomization, and heap isolation.
\section{Discussion}

\toolname may allow a malicious process to compromise the consumer residents in its memory. To this end, we adopt a comprehensive solution that combines address randomization, dedicated heap, MPK protection, and ensuring the atomic execution of the consumer, as discussed in \S\ref{sec:consumer}.  Thus, although the consumer shares the same memory space as user applications, they are still protected. We notice that MPK is available in most of the latest Intel server and client-side CPUs. ARM, AMD, RISC-V, PowerPC, and Itanium CPUs ~\cite{mpk1,mpk2,mpk3,mpk4,vahldiek2019erim} also have similar mechanisms.

Although \toolname prevents attackers from slowing down other applications,  the attacker can still slow down a process by injecting the super producer logic into the process directly. We consider the thread model of this attack too strong for \toolname. Indeed, as long as the attacker can compromise a process, it is straightforward to slow down the process. How to protect a running process from hijacking is beyond the scope of this paper.  

Windows provides the ETW~\cite{etw} framework for provenance collection, but it only has a kernel module and leaves the user-space logic for customization. Thus, we cannot find an ``official'' user-space component for ETW. Nevertheless, the super producer vulnerability is about process scheduling and isolation, which is general to both Linux and Windows. \vspace{-2ex}

\section{Conclusion}

This paper is the first to identify the super producer threat to existing auditing frameworks. Through thorough experiments and case studies, we find attackers can either disable existing auditing frameworks or paralyze the whole system with a super producer. Based on our discovery, we propose a novel auditing framework, \toolname, that addresses the super producer threat by providing resource isolation. Our evaluation shows that \toolname prevents the super producer threat while introducing 6.30\% lower application overhead on average across eight different hardware configurations than Sysdig. 

\section{Acknowledgement}
We sincerely thank our Shepherd and all the anonymous reviewers for their valuable comments. This work was partly supported by the National Key Research and Development Program (No. 2022YFB4501802), the National Natural Science Foundation of China (No. 62172009, No. 62141208) and Huawei Research Fund. 

\bibliographystyle{plain}
\bibliography{main}
\clearpage
\end{document}